\documentstyle[manuscript,prd,aps,tighten]{revtex}
\begin{document}
\draft
\preprint{OKHEP-98-12}
\title{Can the QCD Effective Charge Be Symmetrical  in the
Euclidean and Minkowskian Regions?}

  \author{
K.A. Milton$^{a}$\thanks{E-mail: milton@mail.nhn.ou.edu} and
I.L. Solovtsov$^{b}$\thanks{E-mail: solovtso@thsun1.jinr.ru}
  }
  \address{ $^a$Department of Physics and Astronomy, University of
Oklahoma, Norman, OK 73019 USA\\
$^b$Bogoliubov Laboratory of Theoretical Physics, Joint
Institute for Nuclear Research,
Dubna, Moscow Region, 141980 Russia}

\date{\today}
\maketitle

\begin{abstract}
We study a possible symmetrical behavior of the effective charges defined in
the Euclidean and Minkowskian regions and prove that such symmetry is
inconsistent with the causality principle.
\end{abstract}

A possible way to resolve the ghost-pole problem for the running coupling
constant obtained by using the renormalization group resummation can be
found by imposing  K\"all\'en--Lehmann analyticity, which reflects 
a fundamental property
of local quantum field theory---the principle of causality. This idea
in the QCD case has been elaborated in~\cite{DVS}. The correct analytic
properties of the running coupling give the possibility of a self-consistent
definition of the effective running coupling in the timelike, Minkowskian
region~\cite{MS1,MS98}. In this note we study a possible symmetrical
behavior of the effective charges defined in the spacelike and timelike
domains.

The conventional renormalization group method determines the
running coupling in the Euclidean region. To find a QCD parametrization of
processes that are characterized by  timelike momenta, such as the process
of $e^+e^-$ annihilation into hadrons, one has to use some special procedure
of `analytic continuation' from the Euclidean to the Minkowskian region. To this
end let consider the Adler $D$-function, which corresponds to the vector quark
currents. The perturbative expansion improved by the renormalization group
method in the massless case has the form ($a=\alpha/4\pi$)
\begin{equation}\label{D-pert}
D(Q^2)\propto 1+d_1\bar{a}(Q^2)+d_2\bar{a}^2(Q^2)+\cdots\, .
\end{equation}
The $D$-function is an analytic function in the complex $Q^2$ plane with a cut
along the negative real axes. Defining the effective charge by
\begin{equation}\label{a_t_eff-def}
D(Q^2)\propto 1+d_1\bar{a}^{\rm eff}(Q^2)
\end{equation}
one can see that it must possess the following spectral representation
\begin{equation}\label{a_eff-dispers}
\bar{a}^{\rm eff}(Q^2)=\frac{1}{\pi}\int_0^\infty \frac{d\sigma}{\sigma+Q^2}
\rho(\sigma)\, .
\end{equation}

The appropriate quantity to define the effective charge in the Minkowskian
region (`$s$-channel') is the $R$-ratio for the process of $e^+e^-$
annihilation into hadrons. The structure of the perturbative expansion for
this quantity is similar to the perturbative representation for the Adler
function in the Euclidean region (`$t$-channel') defined by
Eq.~(\ref{D-pert}). The functions $D(Q^2)$ and $R(s)$ in some sense can be
called as `$t$-$s$ dual' functions and, similarly to Eq.~(\ref{a_t_eff-def}), 
one can define the effective charge in the timelike region by
\begin{equation}\label{a_s_eff-def}
R(s)\propto 1+r_1\bar{a}^{\rm eff}_s(s)\, ,
\end{equation}
where the subscript `$s$' means `$s$-channel'.

There is the following connection between these effective charges in the
spacelike and timelike regions:
\begin{equation}
\label{aeff}
\bar{a}^{\rm eff}(Q^2)=Q^2\int_0^{\infty}
\frac{ds}{{(s+Q^2)}^2}\bar{a}_s^{\rm eff}(s)
\end{equation}
and
\begin{equation}
\label{aseff}
\bar{a}_s^{\rm eff}(s)=-\frac{1}{2\pi {\rm i}}
\int _{s-{\rm i}\epsilon} ^{s+{\rm i}\epsilon} \frac{dz}{z}
\bar{a}^{\rm eff}(-z)\, .
\end{equation}
The contour of integration in (\ref{aseff}) lies in the region of the
analyticity of the corresponding integrand.

The effective charge in the $t$-channel is defined through the spectral
function $\rho(\sigma)$ by Eq.~(\ref{a_eff-dispers}). The corresponding
expression for the $s$-channel charge can be written down as follows
\begin{equation}
\label{s-chan-rho}
\bar{a}_s^{\rm eff}(s)=\frac{1}{\pi }\,
\int_s^\infty\,\frac{d\sigma}{\sigma}\,\rho(\sigma)\, .
\end{equation}

Nearly a quarter century ago, Schwinger proposed \cite{schwinger} that the
Gell-Mann--Low function, or the $\beta$-function, in QED could be
represented by a spectral function for the photon propagator, which has a
direct physical meaning. The $\beta$-function of the $s$-channel effective
coupling constant~(\ref{s-chan-rho}) is indeed proportional to the spectral
density, according to Schwinger's identification
\begin{equation}
\label{beta-s}
\beta_s(s)=s\frac{d\bar{a}_s^{\rm eff}(s)}{d s}=-\frac{\varrho(s)}{\pi}.
\end{equation}

Defining the $\beta$-function of the $t$-channel
charge~(\ref{a_eff-dispers})
$
\beta(Q^2)=Q^2{d \bar{a}^{\rm eff}(Q^2)}/{d Q^2}
$
we can write down the following relation between the two $\beta$-functions
\begin{equation}
\label{beta-t-s}
\beta(Q^2)=Q^2\,\int_0^{\infty}\, \frac{d s}{(s+Q^2)^2}\,\beta_s(s)\, .
\end{equation}
Thus, the general properties of the theory lead to the following properties
of the $\beta$-function considered as a function of the Euclidean momentum
$Q^2$: $\beta(Q^2)$ is an analytic function in the complex $Q^2$-plane with
a cut along the negative real axis.

Note that we have defined the effective charge in the $s$-channel. However,
an analogous analysis can be performed for the running coupling and
similar relations and conclusions can be obtained for it as well. In the
framework of perturbation theory, the difference between the $t$- and
$s$-channel running coupling constants appears starting from the three-loop
level (these are the well-known $\pi^2$-terms). Therefore,
$\beta(Q^2)=\beta_s(s=Q^2)+O(3\mbox{-loop})$.  It is also at the three-loop 
level that the $\beta$-function coefficients become renormalization-scheme
dependent.

It is interesting to consider whether there exists a possible solution,
which can be called an $s$-$t$ `self-dual solution' of Eq.~(\ref{beta-t-s}),
in which there is a symmetrical behavior of the charges for the $t$- and
$s$- channels. In this case $\beta(Q^2)=\beta_s(s=Q^2)$ and we have the
following integral equation
\begin{equation}
\label{t-s-dual}
\beta(Q^2)=Q^2\,\int_0^{\infty}\, \frac{d s}{(s+Q^2)^2}\,\beta(s)\, .
\end{equation}

It is clear that there is a ``trivial" solution to Eq.~(\ref{t-s-dual}),
$\beta(Q^2)=\mbox{const}.$ Are there any other solutions? Introduce the
variables $Q^2/\Lambda^2=\exp(x)$ and $s/\Lambda^2=\exp(y)$, and put
$\phi(x)=\beta(Q^2)$ and $\phi(y)=\beta(s)$, so that from
Eq.~(\ref{t-s-dual}), we obtain the integral equation
\begin{equation}
\label{int-eq}
\phi(x)\,=\,\int_{-\infty}^{\infty}\,dy\,K(x-y)\,\phi(y),
\end{equation}
with the kernel
\begin{equation}
\label{k-funct}
K(x)\,=\,\frac{1}{4}\,\frac{1}{\cosh^2(x/2)}\, .
\end{equation}
By applying the Fourier transform to Eq.~(\ref{int-eq}) one finds
\begin{equation}
\label{eq-f}
\tilde{\phi}(p)\,=\,\tilde{K}(p)\,\tilde{\phi}(p)\, ,
\end{equation}
where
\begin{equation}
\label{k-f-explicit}
\tilde{K}(p)\,=\,\frac{\pi\, p}{\sinh(\pi\, p)}\, .
\end{equation}
Possible nontrivial solutions of Eq.~(\ref{eq-f}) appear at the points
for which $\tilde{K}(p)=1$. However, there is only one point of that
sort: $p=0$. Therefore,
\begin{equation}
\label{solution-f}
\tilde{\phi}(p)\,=\,\mbox{const}\cdot \delta(p)\, ,
\end{equation}
which leads to the ``trivial" solution $\beta(Q^2)=\mbox{const}$ and
other $s$-$t$ self-dual solutions are absent.

Thus, behaviors of the running couplings in the spacelike and timelike
regions cannot be symmetrical in any renormalization scheme. It should be
stressed that to reach this conclusion we used only the properties of
analyticity, which reflect the general principle of causality, and,
therefore, this result can be considered as a rigorous consequence of the
first principles of quantum field theory.

\section*{Acknowledgments}

The authors would like to thank D.V.~Shirkov and O.P.~Solovtsova for
interest in this work. Partial support of the work by the US National
Science Foundation, grant PHY-9600421,  by the US Department of Energy,
grant DE-FG-03-98ER41066, and by the RFBR, grant 96-02-16126, is gratefully
acknowledged. I.S. also thanks the high energy group of the
University of Oklahoma for its warm hospitality.


\end{document}